\documentclass{ectj}
\usepackage{subfig}
\usepackage{lipsum}
\usepackage{graphicx,epstopdf}
\DeclareGraphicsExtensions{.ps}
\usepackage[colorlinks = true,citecolor=Blue,urlcolor=Mahogany,linkcolor=Mahogany,]{hyperref}
\usepackage{amsfonts,amssymb,graphics,epsfig,verbatim,bm,latexsym,amsmath,url,amsbsy}
\usepackage[dvipsnames]{xcolor}
\usepackage[top=2.5cm,bottom=2.5cm,left=2.5cm,right=2.5cm,asymmetric]{geometry}
\usepackage[intoc]{nomencl}
\usepackage{geometry}
\usepackage{pdflscape}
\usepackage{booktabs}
\usepackage[font=small,labelfont=bf,
   justification=justified,
   format=plain]{caption}
\makenomenclature
\makeatletter

\newcommand{\fakecaption}{%
  \vskip0.5\baselineskip
  \refstepcounter{table}%
}

\renewcommand{\thesection}{\arabic{section}}
\renewcommand{\theequation}{\arabic{section}.\arabic{equation}}

    \graphicspath{{../figures/}}

  \received{25th May, 2020}
  \volume{1}
  \makeindex

\title[]{Social network-based distancing strategies to flatten the COVID-19 curve in a post-lockdown world}

\author[]{Per Block$^{\ddagger}$, Marion Hoffman$^{\dagger}$, Isabel J. Raabe$^{\ast}$, Jennifer Beam Dowd$^{\ddagger}$}
\author[]{Charles Rahal$^{\ddagger, \mathsection}$, Ridhi Kashyap$^{\ddagger, \mathsection, \mathparagraph}$, Melinda C. Mills$^{\ddagger, \mathsection}$}

\address{$^{\ddagger}$Leverhulme Centre for Demographic Science, Department of Sociology, University of Oxford\\
$^{\dagger}$Department of Humanities, Social and Political Sciences, ETH Zurich\\
$^{*}$Institute of Sociology, University of Zurich\\
$^{\mathsection}$Nuffield College, University of Oxford, Oxford, UK\\
$^{\mathparagraph}$School of Anthropology and Museum Ethnography, University of Oxford, Oxford, UK
\vspace{.1in}}

\def\AmSTeX{$\cal A$\kern-.1667em\lower.5ex\hbox{$\cal M$}\kern-.125em
            $\cal S$-\TeX}
  
\begin{document}

  \begin{abstract}\let\thefootnote\relax\footnotetext{For correspondence: Per Block and Melinda C. Mills, Leverhulme Centre for Demographic Science, Department of Sociology, University of Oxford, OX1 1JD, United Kingdom. Tel: 01865 286170. Email: per.block@sociology.ox.ac.uk and melinda.mills@nuffield.ox.ac.uk. The replication files for this paper including customised functions in the statistics environment R and an example script are available on Zenodo, a general-purpose open-access repository developed under the European OpenAIRE program and operated by CERN (https://zenodo.org/record/3782465).} Social distancing and isolation have been introduced widely to counter the COVID-19 pandemic. However, more moderate contact reduction policies become desirable owing to adverse social, psychological, and economic consequences of a complete or near-complete lockdown. Adopting a social network approach, we evaluate the effectiveness of three targeted distancing strategies designed to `keep the curve flat' and aid compliance in a post-lockdown world. These are limiting interaction to a few repeated contacts, seeking similarity across contacts, and strengthening communities via triadic strategies. We simulate stochastic infection curves that incorporate core elements from infection models, ideal-type social network models, and statistical relational event models. We demonstrate that strategic reduction of contact can strongly increase the efficiency of social distancing measures, introducing the possibility of allowing some social contact while keeping risks low. This approach provides nuanced insights to policy makers for effective social distancing that can mitigate negative consequences of social isolation.
  \end{abstract}

\renewcommand*{\thefootnote}{\arabic{footnote}}

\section{Introduction}\label{introduction}

The non-pharmaceutical intervention of ‘social distancing’ is a central policy to reduce the spread of COVID-19, largely by maintaining physical distance and reducing social interactions \citep{glassetal2006}. The aim is to slow transmission and the growth rate of infections to avoid overburdening health-care systems, widely known as `flattening the curve' \citep{roberts2020}. Social distancing includes bans on public events, the closure of schools, universities and non-essential workplaces, limiting public transportation, travel and movement restrictions, and urging citizens to limit social interactions.

The majority of existing research on mitigating influenza pandemics focus on the effectiveness of different individual measures, such as travel restrictions, school closures, or vaccines (\citealp{fergusuonetal2006,germannetal2006}). Few have simultaneously considered interventions and the structure of social networks. When social networks are examined, it is generally in relation to vaccination \citep{VENTRESCA201375}, contact tracing, or analysing the spread of the virus (\citealp{sunetal2020,wuetal2020}). We outline key behavioural strategies for selective contact reduction that every individual and organisation can adopt to maximise the benefits of limiting contact and engaging in strategic social distancing. Applying insights from social and statistical network science, we demonstrate how changing network configurations of individuals’ contact choices and organisational routines can alter the rate and spread of the virus, by providing guidelines to differentiate between `high-impact' and `low-impact' contacts for disease spread. This can contribute to balancing public health concerns and socio-economic needs for interpersonal interaction. We introduce and assess three strategies: contact with similar people, strengthening contact in communities, and repeatedly interacting with the same people.

Conclusions regarding the effectiveness of non-pharmaceutical public health interventions have often been made on the basis of on ‘expert recommendations’ rather than scientific evidence \citep{bell2006}. During previous outbreaks (e.g. SARS-CoV), social distancing measures such as workplace closures, limiting public gatherings, and travel restrictions were implemented. Cancelling public gatherings and long-distance travel restrictions appears to decrease transmission and morbidity rates \citep{aledortetal2007}. There is mixed evidence regarding the effectiveness of school closures on respiratory infections, possibly because of the timing of school closures, or since this affects only on school-aged children \citep{jacksonetal2013}.

There has been considerably less research on the effectiveness of other types of social distancing measures, such as strategies based on individual's knowledge of their social surrounding. Existing research has demonstrated that interventions are only effective and feasible when the public deems them acceptable \citep{aledortetal2007}. Our approach recognises the social, psychological, and economic cost of -- and potential compliance fatigue with -- complete isolation \citep{morsetal2006}. Fully quarantining non-infected, psychologically vulnerable individuals over prolonged periods can have severe mental health consequences. Many facets of economic and social life require some amount of person-to-person contact. Compliance with recommendations to strategically reduce contact is more favourable than compliance with complete isolation and, thus, can keep the curve flat in the long run. We therefore propose a novel approach that assesses the effectiveness of network adaptations that rely on less confinement and allow some degree of social contact while still `flattening the curve'.

Flattening the (infection) curve represents a decrease in the number of infected individuals at the height of the epidemic, with the incidence of cases distributed over a longer time horizon \citep{roberts2020}. This is largely achieved by reducing the reproduction number (R), which is how many individuals are infected by each carrier. Social distancing policies are implicitly designed to achieve this by limiting the amount of social contact between individuals. By introducing a social network approach, we propose that a decrease in R can simultaneously be achieved by managing the network structure of interpersonal contact. From a social network perspective, the shape of the infection curve is closely related to the concept of network distance or path lengths \citep{wasserman1994social}, which indicates the number of network steps needed to connect two nodes. Popularised examples of network distance include the ‘six degrees of separation’ phenomenon \citep{milgram1967small}, which posits that any two people are connected through at most five acquaintances.

\begin{figure}[!t]
\centering
\includegraphics[width=\textwidth]{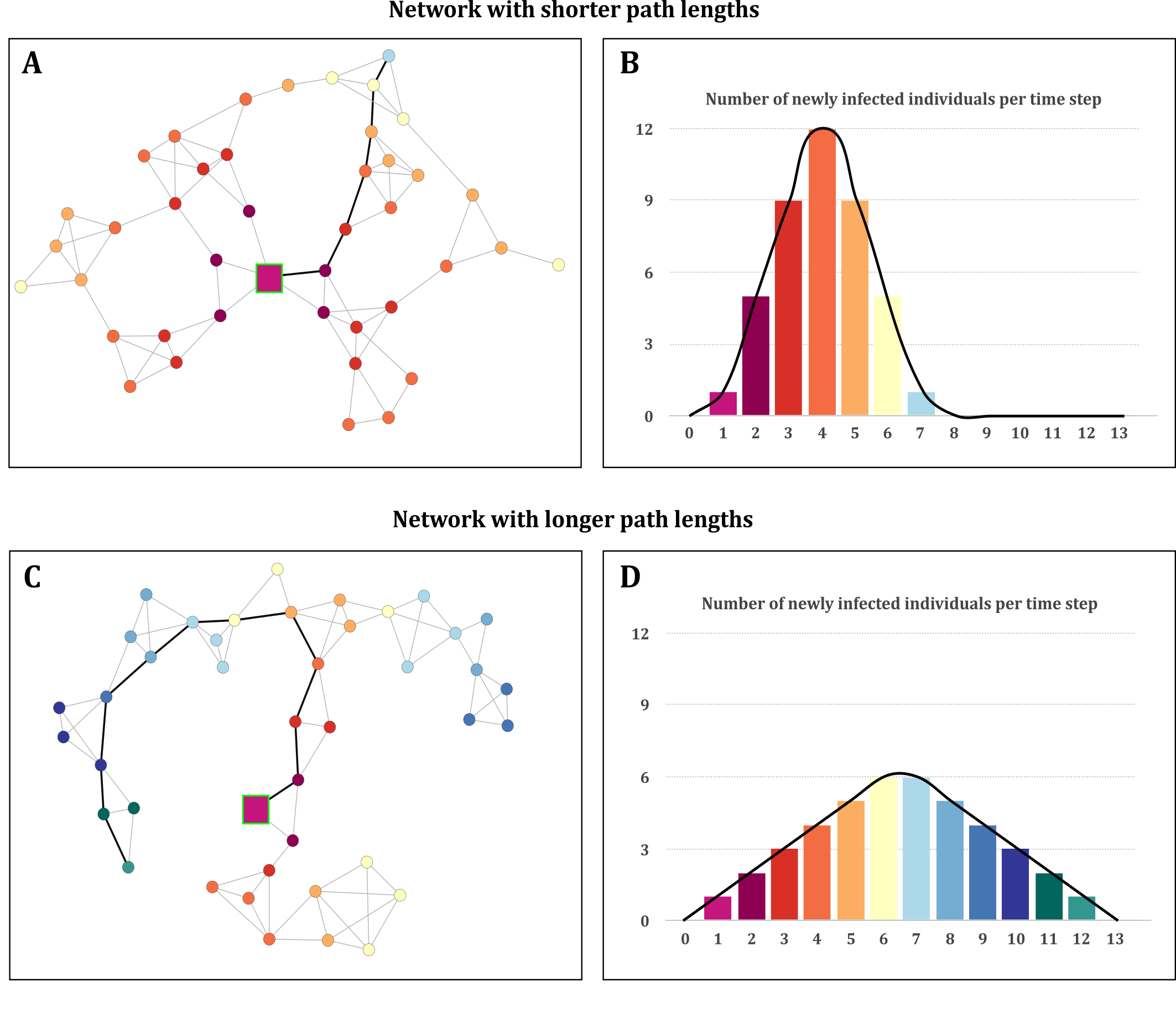}
\caption{\textbf{Two example networks A and C.} Both networks have the same number of nodes (individuals) and ties (social interactions) but different structures which imply different infection curves (B and D). Bold ties highlight the shortest infection path from the infection source to the last infected individual in the respective networks. Network node colour indicates at which step a node is infected and maps onto colours of histogram bars.}
\label{figure1}
\end{figure}

The relation between infection curves and network distance can be illustrated with a simple network infection model (Figure \ref{figure1}). Panels A and C depict two networks with different path lengths, each with one hypothetically infected COVID-19 seed node (purple square). At each time step, the disease spreads from infected nodes to every node to which they are connected; thus, in the first step the disease spreads from the seed node to its direct neighbours. In the second step, it spreads to their neighbours, who are at network distance 2 from the seed node, and so forth. Over time, the virus moves along network ties until all nodes are infected. The example shows that the network distance of a node from the infection source (indicated by node colour in Figure \ref{figure1} A and C) is identical to the number of time-steps until the virus reaches it. The distribution of network distances to the source thus directly maps onto the curve of new infections (Figure \ref{figure1} B and D).

In our example, both networks have the same number of nodes (individuals) and edges (interactions); however, the network depicted in panel C has a much flatter infection curve than the network in panel A, even though all nodes are eventually infected in both cases. This is because the latter network has longer path lengths than the former one -- or in other words -- more network distance between the individuals due to a differing structure of interaction, despite the same absolute contact prevalence. Thus, when adopting a network perspective, flattening the curve is equivalent to increasing the path length from an infected individual to all others, which can be achieved by restructuring contact (besides the generally proposed reduction of contact). Consequently, one aim of social distancing should be increasing the average network distance between individuals by smartly manipulating the structure of interactions. Our illustration shows a viable path to keep the COVID-19 curve flat while allowing some social interaction: we must devise interaction strategies that make real-life networks look more like network C, and less like network A.

We propose a series of strategies for how individuals can make local decisions to achieve this goal. Understanding which types of strategies of targeted contact reduction and social distancing are more efficient in increasing path lengths and flattening the curve can inform how to shift from short-term (complete lockdown) to long-term management of COVID-19 contagion processes. The contact reduction strategies we propose are based on insights of how items flow through networks, such as diseases, memes, information, or ideas (\citealp{wattsetal2006,podolny2001,borgatti2005,centola2010}). Such spread is generally hampered when networks consist of densely connected groups with few connections in-between, such as individuals who live in isolated villages scattered over sparse rural areas \citep{watts1999}. In contrast, contacts that bridge large distances are related to short paths and rapid spread. When commuters travel between these isolated villages, for instance, network distances decrease substantially (\citealp{milgram1967small, centola2010}). Using this knowledge, we can avoid rapid contagion by encouraging social distancing strategies that increase clustering and reduce network short-cuts to reap the largest benefit of reducing social contact and limiting disease spread to a minimum. We propose three strategies aimed at increasing network clustering and eliminating short-cuts.

\begin{figure}[!t]
\centering
\includegraphics[width=\textwidth]{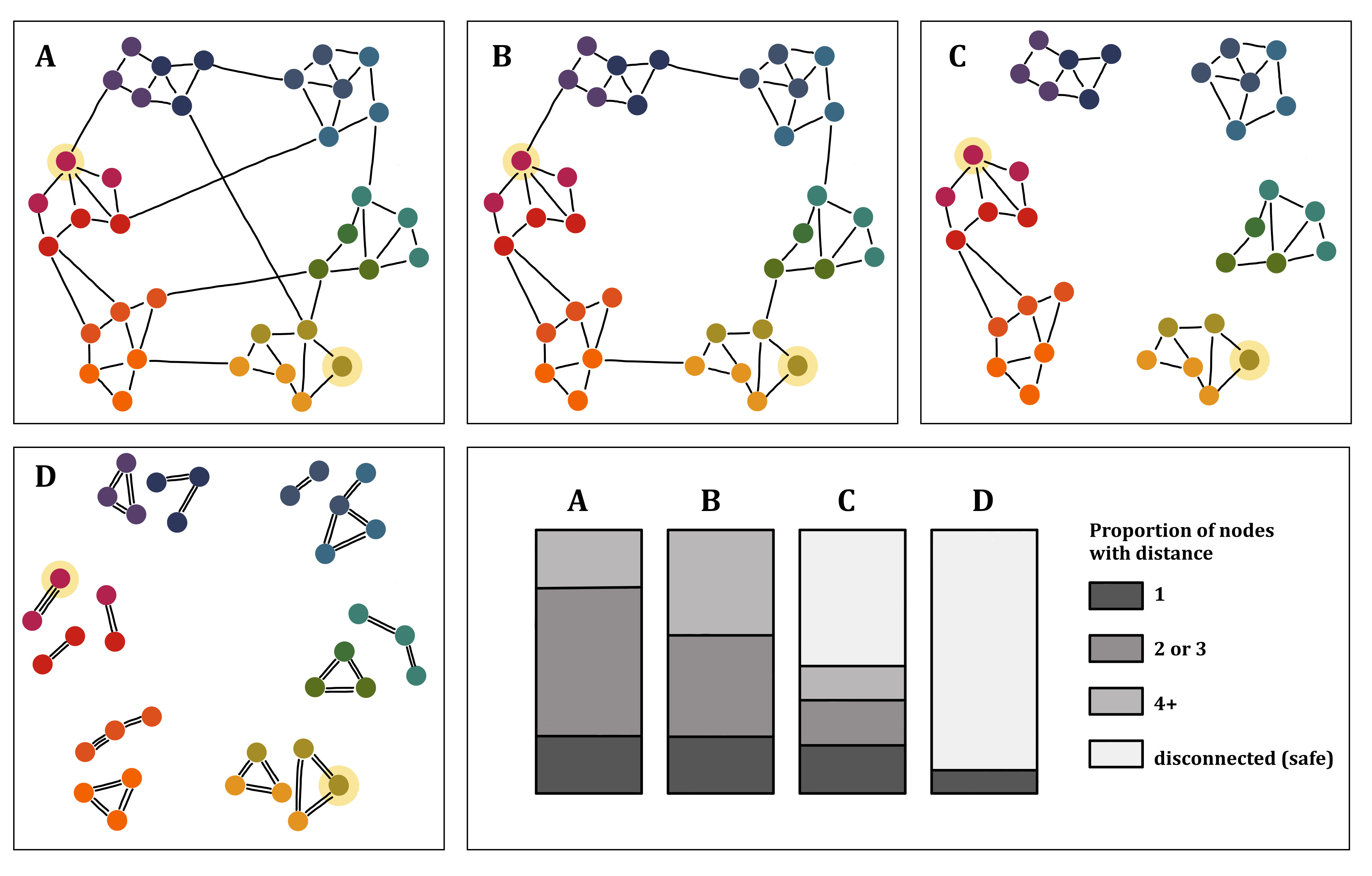}
\caption{\textbf{Example networks that result from the successive tie reduction strategies}. Node colour represents an individual characteristic, where similarity in node colour represents similarity in this characteristic. Node placement represents geographic location of residence. A: initial small world network; B: removing ties to dissimilar others that live far away; C: removing non-embedded ties that are not part of triads or 4-cycles; D: repeating rather than extending contact. Bar graphs show network distances from the infection sources, highlighted in yellow, for the different scenarios.}\label{figure2}
\end{figure}

While more realistic examples of the proposed strategies are simulated in the next section, we first outline the underlying principles of the model in Figure \ref{figure2}. Panel A depicts a network in which densely connected communities are bridged by random, long-range ties. This type of network is commonly known as a `small world network' \citep{centola2010}. It is widely used in simulations, as it represents core features of real-world contact networks, in particular social clustering combined with short network distances, making it particularly useful for our illustration \citep{milgram1967small}. Within clusters, individuals are similar to each other, indicated by their node colour, and live in the same neighbourhood, indicated by node location. The further away two clusters are in the figure, the further they live from each other and the more dissimilar their members. Panels A to D illustrate the successive, targeted contact reduction strategies, while the bar-graph depicts the distribution of distances of all individuals from one of the two highlighted infection sources.

\textbf{Strategy 1}: `Seek similarity' strategy: Reduce geographic and socio-demographic difference to contact partners (A to B in Fig. \ref{figure2}). In the first strategy, individuals choose their contact partners based on their individual characteristics. Generally, individuals tend to have contact others who share common attributes, such as those in the same neighbourhood (geographical), or of similar income or socio-demographic characteristics such as age (\citealp{feld1981,rivera2010,mcphersonetal2001}). The tendency to interact with similar others is called ‘homophily’ in the sociological network literature \citep{rivera2010} and is a ubiquitous and well-established feature of social networks (thus, we use ‘seek similarity’ strategy and ‘homophily’ strategy interchangeably). Because we are mostly connected to similar others, contact with dissimilar individuals tends to bridge to more distant communities. Restricting one’s contact to those most similar helps limit network bridges that substantially reduce network path lengths. This entails choosing to interact with those geographically proximate (e.g., living in the same neighbourhood), or individuals with similar characteristics (e.g., age). Panel B in Figure \ref{figure2} shows the network structure after the implementation of this strategy of tie reduction. The associated bar-graph illustrates that following this network-based intervention, a substantial number of nodes are at a larger distance from the infection source. This strategy will be successful when the characteristic or variable which determines the communities can take on a variety of different (categorical or continuous) values for different individuals, thereby promoting the formation of small communities. A broader split, such as along gender or ethnic lines does not promise measurable success but will instead likely exacerbate the negative consequences of distancing measures. This strategy is supported by epidemiological modelling which suggests that co-residence and mixing of individuals from different ages strongly increases the spread of infectious disease, such as COVID-19 \citep{pellisetal2020}. Providing a concrete example, if people only interact with others in a 3-block radius (increase geographic similarity), more than 30 transmission events would be necessary for a virus to travel 100 blocks. Workplaces where many individuals come together could, for instance, implement routines to decrease contact between groups from different geographic areas or age-groups.

\textbf{Strategy 2}: `Strengthen triadic communities' clustering strategy: Increase triadic clustering among contact partners (B to C in Fig. \ref{figure2}). For the second strategy, individuals must consider with whom their contact partners usually interact. A common feature of contact networks is ‘triadic closure’, referring to the fact that contact partners of an individual tend to be connected themselves (\citealp{feld1981,granovetter1973,goodreau2009}). Tie embedding in triads is a particularly useful topology for containing epidemic outbreaks. Consider a closed triad of individuals $i$, $j$, and $h$. When $i$ infects $j$ and $h$, the connection between $j$ and $h$ does not contribute to further disease spread: it is a `redundant' contact \citep{burt1995structural}. When comparing networks with an identical number of connections, networks with more redundant ties tend to have longer path lengths. Accordingly, when removing contact to others, one should prioritize removing ties not embedded in triads, since these ties generally decrease path lengths. In practice, this means that physical contact should be curtailed with people who are not also connected to one’s usual other social contacts. Panel C in Figure \ref{figure2} illustrates the structure if ties that are not part of closed triads or 4-cycles are removed. In this ideal-type example, this intervention not only further reduces the network distance of many nodes from the infection sources, but also creates isolated communities or that cannot be infected by the virus.

\textbf{Strategy 3}: `Repeat contact and build micro-communities' strategy: Repeated contact to same others, rather than changing interaction partners (C to D in Fig. \ref{figure2}). For the third strategy, individuals need to consider who they want to regularly interact with and, over time, restrict interaction to those people; this reduces the number of contact partners rather than number of interactions, which is particularly important when contact is necessary for psychological well-being. This strategy of limiting contact to very few others with repeated interactions is in the spirit of a social contract with others to create micro-communities to only interact within the same group delineated by common agreement. Although this requires coordination, micro-communities would be difficult for a virus to penetrate, or -- importantly -- if the infection is contracted by one contact, for the virus to spread further. Another implication of this strategy includes the repetition of interaction with others that overlap across more than one contact group. For example, meeting co-workers outside of work for socializing will have less of an impact on the virus spread relative to a separate group of friends, since a potential infection path already exists. Having tight and consistent networks of medical or community-based carers for those more vulnerable to COVID-19 (elderly, pre-existing conditions) limits the transmission chain. Organisations can leverage this strategy by structuring staggered and grouped shifts so that individuals have repeated physical contact with a limited group rather than dispersing throughout an organisation. Panel D in Figure \ref{figure2} illustrates the resulting network structure.

Strategy 2 and 3 are similar in that they build on pre-existing network structures. However, their difference lies in the determinants of individual interaction. Strategy 2 relies on a stable and established network structure of durable relations: who are members of my usual `groups' (e.g., friends, family, co-workers) and which pairs of individuals among my usual contacts interacts with each other, too? Strategy 3 relies on a strategic decision to form most convenient and effective ``interaction bubbles'' and repeat contact to them over time. In this sense, strategy 2 is easier to implement, since individuals are able to shape their contacts themselves,  while strategy 3 requires coordinated action of everyone involved in a given ``bubble''. Until now, we have illustrated our strategies with an intuitive but stylized model of epidemic spread. We now demonstrate how our three contact strategies impact infection curves using more formal stochastic infection models that incorporate core elements from infection models, ideal-type network models and statistical relational event models. These strategies are compared to a baseline (null) model that represents how the COVID-19 infection would spread if there was unrestricted contact (i.e., no social distancing). 

First, our model draws from classical disease modelling (\citealp{kermack1927contribution,anderson1992infectious}), in which individuals (actors) can be in four states: susceptible, exposed (infected but not yet infectious), infectious, and recovered (no longer susceptible to infection). Most actors begin in the susceptible state, while $q$ random actors are in the infectious state (one per thousand in our simulations). This can represent, for example, the post-lockdown scenario in which only a few cases of COVID-19 remain in the population; however, variation of $q$ might also be used to determine the levels at which a lock-down can be eased. During the simulation, susceptible actors can transition to the exposed state by having contact with infectious others (contact partners will be called `alters' from here on). Whether contact between a susceptible actor and an infectious alter results in contagion is determined probabilistically. A designated time after becoming exposed, actors become infectious themselves, and later move to the recovered state after another fixed amount of time.

Second, as in many previous modelling efforts of the dynamics of epidemics such as influenza, we do not assume homogeneous contact probabilities in an affected population but rather impose a network structure that limits contact opportunities between actors (\citealp{newman2002,halloranetal2008,salatheetal2010}). This network represents the typical contact people had in a pre-COVID-19 world. The networks we generate stochastically for our model follow fairly standard ideal-type network generating approaches. Representing place of residence, actors are assumed to have a geographic location, determined by coordinates in a two-dimensional space. They are members of groups, such as households, institutions like schools or workplaces, and have individual attributes, such as age, education, or income. Network ties are generated so that actors have some connections to geographically close alters, some ties to members of the same groups (representing e.g., co-workers), some ties to alters with similar attributes (e.g., similar age), and, finally, some ties to random alters in the population. The generated networks represent the structure of alters that an actor can possibly interact with. They represent the members of their so-called `social circles' (\citealp{watts1999,feld1981,block2018}) with whom they interact in their normal, pre-COVID life (including family, friends, schoolmates or co-workers). The exact algorithms which define the networks are described in the Methods section.

In the third component of the model, actors in the network interact at discrete times with alters with which they have a connection in the underlying network, or in other words, someone they meet from their usual social contacts. This represents the actual contact people have in their lives during which the disease can be transmitted from infectious actors to susceptible alters. Notably, in contrast to other modelling approaches, we do not assume that actors interact with alters in their personal network with uniform probability (i.e. at random), but, rather, that they are purposeful actors who make strategic choices about interactions. These strategic choices are at the core of our advice for policy interventions, where individuals can strategically increase the efficiency of social distancing. In our model, all choices are stochastic; strategies increase the likelihood of interacting with specific alters but are not deterministic. The exact formulation of with whom to interact follows a multinomial logit model to choose among possible interaction partners, given by the network structure. This type of model has previously been used in network evolution \citep{snijders2001} and relational event models (\citealp{butts2008,stadtfeld2017}).

Our simulations explore the three interaction strategies we propose. First, in our `seek similarity' strategy, actors choose to interact predominantly with others that are similar to themselves based on one or several specified attributes used at the network generation stage. Second, actors can adopt our `strengthen triadic community' or triadic strategy and choose to mostly interact with alters that have common connections in the underlying network. Third, adopting our `repeat contact' strategy, actors can base their choices on whom they have interacted with in their previous contacts, both as sender and receiver of an interaction. In each case, a separate statistical parameter in the multinomial model determines the probabilities of interaction partners based on the: (i) similarity of alters, (ii) number of common contacts the actor and alter have; and, (iii) repeat interaction with one of the last $j$ contact partners (see Methods). In our analyses, these three strategies are compared to a baseline case that mirrors the simple reduction of contact in which individuals have the same amount of interactions but choose randomly amongst their network contacts (a na\"{i}ve contact reduction strategy) and a null model that represents unbridled contact without any social distancing. To make the comparison of interaction strategies independent of the arbitrary size of statistical parameters, we empirically calibrate parameters so that the average entropy in the probability distribution that represents the likelihood of different interaction choices is identical for all strategies, as documented in our Methods section \citep{snijders2004}.

Following an initial analysis that represents a benchmark scenario of our disease model, we present a series of variations in modelling parameters that explore alternative scenarios and ensure our main conclusions are independent of user-defined parameters and arbitrary modelling choices. Variations are fully described in the methods section and include: (i) different operationalisations of homophily; (ii) the effect of employing mixed strategies; (iii) number of actors in the simulation; (iv) varying the underlying network structure in the simulations; (v) length of the interval in which actors are exposed relative to the time they are infectious; and (vi) the infectiousness of the virus.

\section{Results}

\begin{figure}[!t]
\centering
\includegraphics[width=0.85\textwidth]{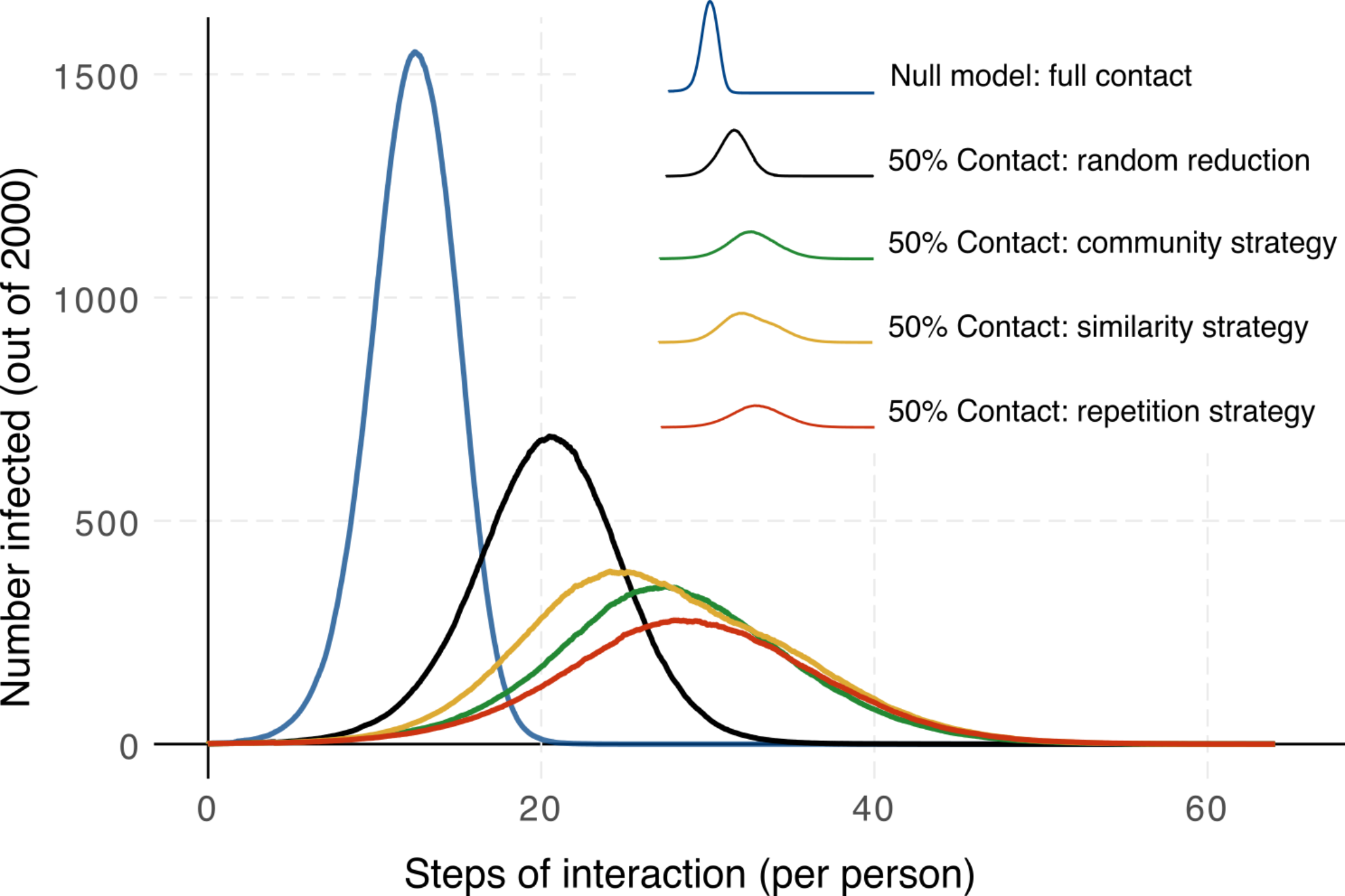}
\caption{\textbf{Average infection curves}. Curves compare 4 contact reduction strategies to the null model of no social distancing. Underlying network structure includes 2000 actors and the benchmark network characteristics described in the main text.}\label{figure3}
\end{figure}

The average outcome of the benchmark scenario is presented in Figure \ref{figure3}. The x-axis represents time as measured in simulation steps per actor and the y-axis the number of individuals infected at this time step out of a total population of 2,000. Curves are averaged over 40 simulation runs. The first scenario in blue shows a null or control interaction model in which there is no social distancing and actors interact at random. The next four strategies all employ a 50\% contact reduction relative to the null model and compare different contact reduction strategies. The black line represents naïve social distancing in which actors reduce contact in a random fashion. The golden line represents the infection curve when actors employ our first `seek similarity' strategy. The green line models our second triadic strategy of `strengthening communities' and represents the associated infection curve. Finally, the dark red line shows how infections develop when actors employ our third strategy of `repeat contact'.

All three of our strategies substantially slow the spread of the virus compared to either no intervention or simple, un-strategic social distancing. The most effective is the strategic reduction of interaction with repeated contacts. In comparison to the random contact reduction strategy, the average infection curve delays the peak of infections by 37\%, decreases the height of the peak by 60\%, and results in 30\% fewer infected individuals at the end of the simulation. This is marginally more efficient than the triadic strategy and the homophily strategy, in this order (delay of peak 18\% and 34\%, decrease in peak height of 44\% and 49\%, and reduction of infected individuals by 2\% and 19\%, for homophily and triadic strategies, respectively). Note that these metrics cannot be interpreted as general estimates of the efficiency of these strategies in real-world networks. Summarizing the sensitivity and robustness analyses carried out, strategic contact reduction has a substantive effect on flattening the curve compared to simple social distancing consistently across all scenarios. However, interesting variations occur as discussed below. Full average infection curves and results description for all model variations are presented in the Supplementary Information.

\subsection*{Different operationalisations of homophily}

In the benchmark model, the ‘seek similarity’ strategy was employed on one demographic attribute. However, in real-world social networks, individuals are homophilous on multiple attributes \citep{blockgrund2014}. Furthermore, the benchmark model only uses demographic homophily, while we previously also discuss the importance of geographic homophily. In a variation of the homophily strategy, we show that using geographic homophily for contact reduction is highly efficient, much more than homophily based on demographic attributes (Figure \ref{figures1}b.). Geographic homophily effectively eliminates contacts to distant others in the network. In a further analysis, we compare the benefits of using one dimension of demographic homophily or a composite of two dimensions that structure the network. This explores whether we should focus on interacting with persons similar in one dedicated dimension or seek out others who are as similar as possible in multiple dimensions. Encouragingly, the focus on one strategic dimension of homophily provides similar outcomes to reducing overall demographic distance, meaning that homophily should be encouraged on the dimension that has the least adverse consequences for societal cohesion. Infection curves are presented in Figure \ref{figures1}c.-d.

\subsection*{Employing mixed strategies}

Since most individuals in a post-lockdown world need to interact across multiple social circles (e.g., workplace, extended family), employing only one strategy might not be practical. A mix of different strategies could therefore be more realistic for everyday use. We tested how four possible combinations of mixing strategies (three two-way combinations and one three-way combination) compare to the single strategies of seeking similarity and strengthening communities. We find that the combined strategies are comparably as effective as single strategies (Figure \ref{figures2}) and can be recommended as alternatives if single strategies are not practicable in some contexts. Importantly, each combination performs better in limiting infection spread compared to the naïve contact reduction strategy.

\subsection*{Varying the number of actors in the simulation}

The computational complexity of our simulation prohibits assessing disease dynamics in very large networks (e.g. 100k+ actors), even on large distributed systems. Nevertheless, we can compare simulations using the same local network topology as the benchmark model on networks of 500, 1000, 2000, and 4000 actors. Reassuringly, we find no variation of the relative effectiveness of the different interaction strategies by network size (see Figure \ref{figures3}). While this does not fully allow extrapolation to very large networks, it provides initial support that disease spread under the model could be similar within differently sized sub-regions of larger, real-world networks.

\subsection*{Varying the underlying network structure}

The generating process of the ideal-type network that provides the opportunity structure among individuals with whom they can interact contains multiple degrees of freedom. These include the average number of contacts and the importance of different foci (geography, groups, and attributes) in structuring contact. We provide infection curves for multiple scenarios in the Supplementary Information (Figure \ref{figures4} and \ref{figures5}), showing that our strategies work mostly independent of the underlying structure. A first noteworthy finding from these simulations is that in networks with fewer connection opportunities, all strategies have much larger benefits compared to networks with more connection opportunities (panels C and D in Figure \ref{figures4}). In fact, the triadic strategy does not seem to work anymore in the scenarios with very high average connectivity in the underlying network – most likely because of a large number of closed triangles. This shows that in communities that have lower connectivity, spread can be contained even better. As a second finding, we see that in the case where the underlying network is not structured by homophily, the homophily strategy does not work (panel C in Figure \ref{figures5}), illustrating how the strategy relies on predetermined structural network features.

\subsection*{Variation in infectiousness and the length of the exposed period}

Average infection curves under conditions of differences in infectiousness of the virus, and variations of the time individuals are in the state “exposed” relative to the time of being in the state “infectious” do not influence the relative effectiveness of the different strategies and are presented in Figures \ref{figures6} and \ref{figures7} respectively.

\section{Discussion and conclusion}

In the absence of a vaccine against COVID-19, governments and organisations face economic and social pressures to gradually and safely open up societies but lack scientific evidence on how to best do so. We provide clear social network-based strategies to empower individuals and organisations to adopt safer contact patterns across multiple domains by enabling individuals to differentiate between `high-impact’ and `low-impact’ contacts. The result may also be higher compliance since actors will hold the power to strategically adjust their interactions without being requested to fully isolate. Instead of blanket self-isolation policies, the emphasis on similar, community-based, and repetitive contacts is both easy to understand and implement thus making distancing measures more palatable over longer periods of time. 

How can this be applied to real-world settings? When a firm lock-down is no longer mandated or recommended, it is likely that individuals will want or need to interact in different social circles, e.g. at the workplace and with the wider family. Consequently, the simple one-at-a-time strategic recommendations we analysed in most simulations might be impossible to follow strictly by some. Our sensitivity analysis using mixed strategies addresses this concern. For example, does mixing the three strategies still provide benefits or do they counteract one another in their effect? Reassuringly, a mix of strategies still provided comparable benefits to single strategies, compared to na\"{i}ve contact reduction. Further modelling is needed to assess the implications in a variety of contexts. However, when approaching this issue from a policy perspective, designing steps to ease lockdown can be done with potential behavioural recommendations in mind: if network structures and demographic characteristics of individuals in particular regions suggest that the use of one strategy will yield the best results, decisions on which contact opportunities to allow -- such as opening schools or local shops -- might be taken so that this strategy can be adhered to most easily.

A second discussion point concerns the potential unintended consequences of recommending our triadic and homophilous strategies. Advocating the creation of small communities and contact to mostly similar others can potentially result in the long-term reduction of intergroup contact and an associated rise in inequality \citep{dimaggio2012}. In our simulations we explored this concern by comparing the scenarios when homophilous ties in the underlying network are formed following similarity in multiple dimensions, e.g. age and income. Our test of whether minimising the overall difference in attributes of contacts versus only reducing homophily on one dimension suggests that choosing one salient attribute can already go a long way. Thus, policymakers can make smart choices in deciding which attribute people should pay attention to, keeping the potential social consequences in mind. Nevertheless, understanding the long-term social consequences of which types of public spaces are opened and, accordingly, which types of interaction are allowed should be a major policy concern.

A number of concrete policy guidelines can be deduced from our network-based strategies. For hospital or essential workers, risk is minimized in sustained shifts with similar composition of employees (i.e., repeating contact) and, to distribute people into shifts based on, for example, residential proximity where possible (i.e., homophily). In workplaces and schools, staggering shifts and lessons with different start, end and break-times by discrete organisational units and classrooms will keep contact in small groups and reduce contact between them. When providing private or home care to the elderly or vulnerable, the same person should visit rather than rotating or taking turns, but that person should be the one with fewest bridging ties to other groups and who lives the closest  (geographically). Repeated social meetings of individuals of similar ages that live alone carry a comparatively low risk. However, in a household of five, when each person interacts with disparate sets of friends, many short cuts are being formed that are potentially connected to a very high risk of spreading the disease.

Simple behavioural rules can go a long way in `keeping the curve flat'. As the pressure grows throughout a pandemic to ease stringent lockdown measures increases to relieve social, psychological, and economic burdens, our approach provides insights to individuals, governments and organisations about three simple strategies: interacting with similar types of people, strengthening interaction within communities, and repeating interaction with the same people. 

\section{Methods}

\subsection*{Generation of stylised networks}

The stylised binary networks $x$ that represent interaction opportunities of individuals are generated as the composite of four sub-processes. Jointly, the sub-processes create networks that have realistic values of local clustering, path-lengths, and homophily. All ties in the network are defined as undirected. The number of actors in the network is denoted by $n$.

The first sub-process represents tie formation based on geographic proximity \citep{hamill2009}. First, all actors in the network are randomly placed into a two-dimensional square. Second, each actor draws the number of contacts which it forms in this subprocess $d_{geo,i}$ from a uniform distribution between $d_{geo,min}$ and $d_{geo,max}$; for example, if $d_{geo,min}=10$ and $d_{geo,max}$=20, every actor forms a random number of ties between 10 and 20 in this sub-process. Third, the user-defined density in geographic tie-formation $g_{geo}$ defines the geographic proximity of contacts drawn, so that actor $i$ randomly forms $d_{geo,i}$ ties among those $\frac{d_{geo,i}}{g_{geo}}$ that are closed in Euclidean distance from actor $i$. For example, if actor $i$ is posed to form $d_{geo,i}$=12 ties and $g_{geo}$=0.5, the actor randomly chooses 12 out of the 24 closest alters to form a tie to. Across all simulated networks we set $g_{geo}$=0.3. Fourth, unilateral choices (where only $i$ selected $j$ but not vice versa) are symmetrised so that a non-directed connection exists between the actors.

The second sub-process represents tie formation in organizational foci, e.g. workplaces \citep{herbert2015}. First, each actor is randomly assigned to a group so that all groups have on average $m$ members. Second, each actor forms ties at random to other members within the same groups with a probability of $g_{groups}$. For example, when $m$=10 and $g_{groups}$=0.5, a tie from each actor to every alter in the same group is formed with a probability of 50\%. Third, unilateral ties are symmetrised as above.
 
The third sub-process represents tie-formation based on homophily, for example similarity in age or income \citep{pellisetal2020}. First, each actor is assigned an individual attribute $a_i$ between 0 and 100 with uniform probability (the scale of $a_i$ cancels later in the model). Second, for each actor, the normalized similarity $sim_{i,j}$ to all alters $j$ is calculated, which is one minus the absolute difference between $a_i$ and $a_j$ for actor $j$, divided by 100 (the range of the variable), so that $sim_{i,j}$=1 in case $i$ and $j$ have the identical value of $a$ and $sim_{i,j}$=0 if they are at opposite ends of the scale. Third, each actor draws the number of contacts it forms in this subprocess $d_{homo,i}$ from a uniform distribution between $d_{homo,min}$ and $d_{homo,max}$. Fourth, each actor creates $d_{homo,i}$ ties to alters $j$ in the networks with a probability that is proportional to (sim$_{i,j}$)$^w$,  where higher values of $w$ mean that individuals prefer more similar others. Across all reported simulations, we set $w$=2. Fifth, unilateral ties are symmetrised as above.

The fourth sub-process represents haphazard ties that are not captured by any of the above processes. Here simply $z$ ties per actor are created with respect to randomly chosen alters.

\subsection*{Definition of simulation model}

Let the binary network $x$ represent the underlying social ties between $n$ individuals, labeled from 1 to $n$. Each node $i$ is characterized by a set of attributes $a_i^k$ (such as age or location). Our model aims to reproduce the process of individuals interacting with some of their social connections. Similar to the classic SIR model \citep{kermack1927contribution} and its SEIR extension \citep{anderson1992infectious}, we assume that individuals can be in four different states: either susceptible to the disease, exposed (infected but not yet infectious), infectious, or recovered. Infection occurs through social interactions, which are modeled in a similar fashion to the Dynamic Actor-Oriented Model \citep{salatheetal2010} developed for relational events. More specifically, our model is comprised of the following steps:

\begin{enumerate}
\item At each step of the process, one individual is picked at random and initiates an interaction with probability $\pi_{contact}$.
\item An actor initiating an interaction can only pick one interaction partner. Only potential partners as defined by the network $x$ can be chosen. The decision to interact is unilateral and depends on characteristics of the two persons through a probability model $p$.
\item An infectious individual infects a healthy person when they interact, who then becomes exposed. This contagion occurs with the probability $\pi_{infection}$.
\item After a fixed number of steps ($T_{exposure}$), an exposed individual becomes infectious.
\item After becoming infectious, recovery occurs within $T_{recovery}$ steps. Once recovered, individuals can no longer be infected.
\item The process ends once there is no longer anyone exposed or infectious.
\end{enumerate}

The steps of the model are illustrated in Figure \ref{figure4}. One can note that the mechanics of the infection align with previously proposed agent-based versions of the SIR and SEIR models (\citealp{chowell2016,siettos2013}). Together, the probabilities $\pi_{contact}$ and $\pi_{infection}$ play a similar role as the classic infectivity rate ($\beta$) in SIR models. The rate  models the average number of contacts per person (modelled here through $\pi_{contact}$) and the likelihood of infection (represented by $\pi_{infection}$), however the equivalence is not direct due to the added step of the interaction probability ($p$). The exposure and recovery times replace the classic exposure and recovery rates (often traditionally denoted as $\sigma$ %check this!!!
 and $y$) in a straightforward manner. Let us turn to the definition of the probability model $p$. Let $N_i$ be the set of potential contacts, or alters, $j$ of a given individual $i$ in the network $x$. We define for each step $t$ of the process, $L_i(j,t)$ %check this
 as the number of prior interactions between $i$ and an alter $j$, within the last $K$ interactions of $i$. In our simulations, the number $K$ was arbitrarily set to $2$ but can be easily adjusted in the replication files. For each alter $j \in N$, the value $s(i,j)$ represents the statistic driving the strategical choice of $i$ to pick $j$. Specifically, we define three different ways depending on whether the homophily, the triadic, or the repetition strategy is chosen (however, arbitrary other statistics can be defined). The statistic $s_{homophily}$ accounts for the level of similarity between $i$ and $j$ given a set of attributes, $s_{triadic}$ corresponds to the number of alters they share, and $s_{repetition}$ is the count of previous interactions within the last $K$ contacts of $i$. In practice, these statistics are calculated as:

\begin{figure}[!t]
\centering
\includegraphics[width=1\textwidth]{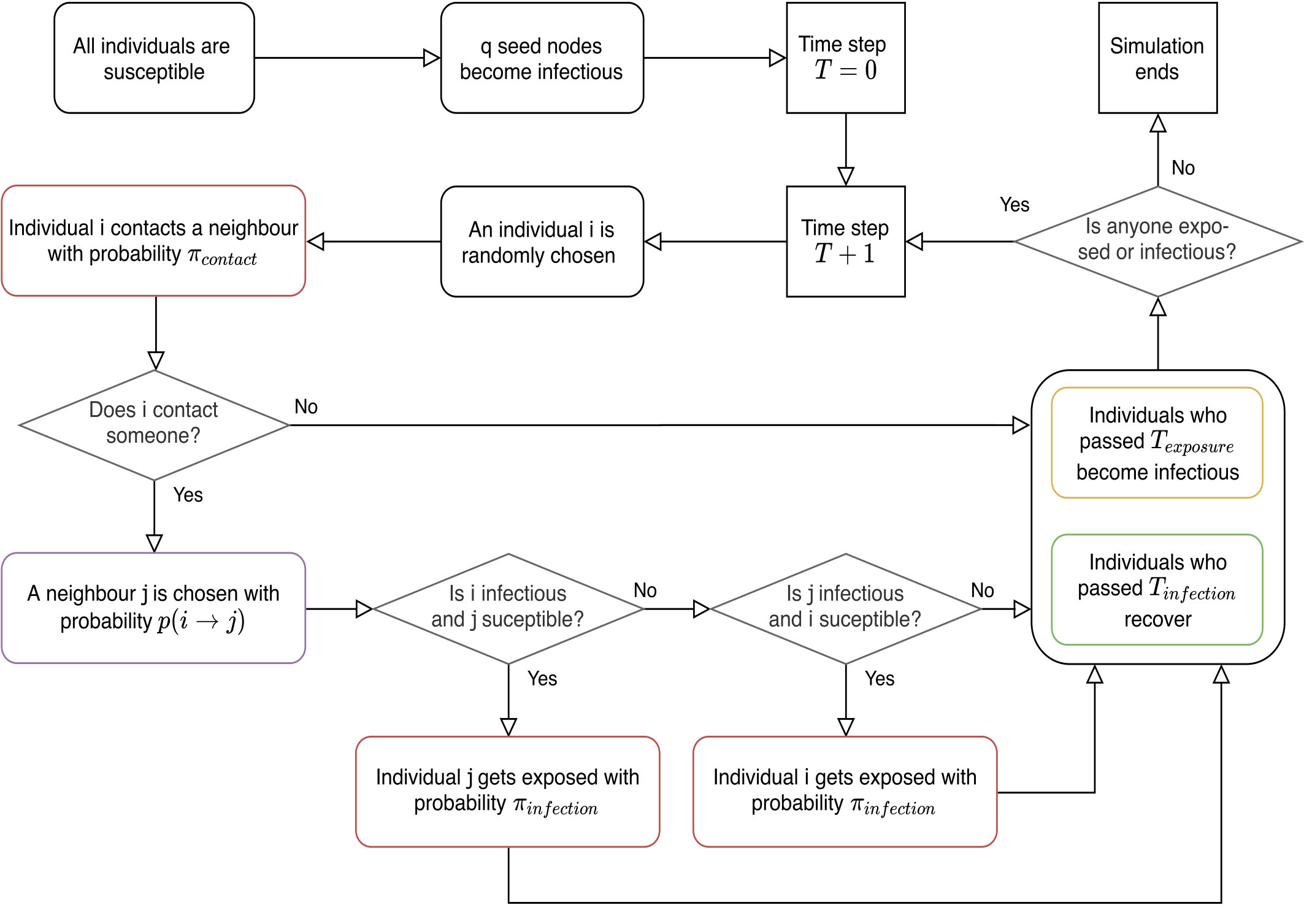}
\caption{Figure 4. Flowchart of the simulation model. Squares indicate updating steps to individuals or the entire system. Diamond shapes represent decisions that determine the subsequent step in the simulation. In the iterative part of the model, a random individual $i$ is chosen, to initiate an interactions with probability $\pi_{contact}$ . In case an interaction is initiated, a contact partner $j$ is chosen with probability $p(i \rightarrow j)$ following a multinomial choice model. If either interaction partner is infectious and the other is susceptible, contagion occurs with probability $\pi_{infection}$ . Subsequently, among all individuals in the simulation, those that are in the exposed state for more than $T_{exposure}$  transition to infectious state and those that are in the infectious state for more than $T_{infection}$ recover. These recursive steps are repeated until all individual are either in the susceptible or recovered state. The colors red, green, and yellow relate closely to the steps in the SEIR model, where red squares govern the transition from susceptible to exposed, the yellow square governs the transitions from exposed to infectious, and the green square governs the transition from infectious to recovered. The purple square represents the step at which individuals strategically chose interaction partners to limit disease spread.}\label{figure4}
\end{figure}

\begin{equation}
s_{homophily} (i,j)= 1- \frac{\sqrt{\sum_k (a_i^k - a_j^k)^2}}{\underset{i,j}\max(\sqrt{\sum_k (a_i^k-a_j^k)^2})-\underset{i,j}\min(\sqrt{\sum_k (a_i^k-a_j^k)^2})}
\end{equation}  

\begin{equation}
s_{triadic} (i,j)=\sum_{k=1}^n x_{i,k} x_{j,k}
\end{equation}
\begin{equation}
s_{repetition} (i,j)=L_i (j,t)
\end{equation}

The probability for $i$ to pick $j$ is defined as a multinomial choice probability \citep{mcfadden1973conditional}, following the logic of previous relational event \citep{block2018} and stochastic network models \citep{halloranetal2008}. The intuition behind this distribution is that each potential partner in $N_i$ is assigned an objective function value, and choosing a partner is based on these values. Mathematically, the objective function is an exponentiated linear function of the statistic $s$($i$,$j$), weighted by a parameter $\alpha$. We further assume that individuals can reduce a certain percentage of their interactions. Considering the probability ($\pi_{contact}$) of initiating an interaction in the first place, the relevant probability distribution becomes:
 \begin{equation}
p(i \rightarrow j | \pi_{contact},\alpha)=\frac{\pi_{contact} exp(\alpha \times s(i,j))}{\sum_{j^{`} \in N_i} exp(\alpha \times s(i,j^{`}))}
\end{equation}

These probabilities can be loosely interpreted in terms of log-odd ratios, similarly to logit models. Given two potential partners $j_1$ and $j_2$ for which the statistic $s$ increases of one unit (i.e. $s$($i$,$j_2$)=$s$($i$,$j_1$)+1), the following log ratio simplifies to:

\begin{equation}
\textrm{log} \frac{p(i \rightarrow j_2 | \pi_{contact},\beta)}{p(i \rightarrow j_1  | \pi_{contact},\beta)}=\alpha
\end{equation}

For example, if we use $s$=$s_{repetition}$ and $\alpha_{repetition}=\textrm{log(2)}$, the probability of picking one alter present in the last contacts of $i$ is twice as high as picking another alter who is not.

\subsection*{Calibration of model parameters}

The strategy of picking a neighbor at random corresponds to the model without any statistic $s$, reducing the probability distribution to a uniform one. For the three other strategies, the parameters $\alpha_{homophily}$, $\alpha_{triadic}$, and $\alpha_{repetition}$ are adjusted to keep the models comparable. To this end, we use the measure of explained variation for dynamic network models devised by Snijders \citep{snijders2004}. This measure builds upon the Shannon entropy and can be applied to our model to assess the degree of certainty in the choices individuals make. For a given individual $i$ at a step $t$, this measure is defined as:

\begin{equation}
r_H (i,t | \pi_{contact} , \alpha) = 1 + \frac{\sum_{j \in N_i} p(i \rightarrow j | \pi_{contact}, \alpha) \times log_2 (p(i \rightarrow j | \pi_{contact}, \alpha))}{log_2(|N_i|)}
\end{equation}

Intuitively, this measure equals 0 in the case of the random strategy where the probability of picking any alter is identical. It increases whenever some outcomes are favored over others and equals 1 if one outcome has all of the probability mass. Since the model assumes all individuals are equally likely to initiate interactions, we can average this measure over all actors. Moreover, in the case of the repetition strategy, the measure is time dependent. In that case, we use its expected value over the whole process. We finally use the following aggregated measure in order to evaluate the certainty of outcomes of a specific strategy:

\begin{equation}
R_H (\pi_{contact}, \alpha)=\frac{1}{n}\sum_{i=1}^n E[r_H (i,t)]
\end{equation}

For this article, we first fix the parameter $\alpha_{repetition}$ at a value of 2.5, and calculate an estimated value $\hat{R}_H$($\pi_{contact}$, $\alpha_{repetition}$) of this measure. This experience-based parameter choice results in an associated $R_H$ value between 0.3 and 0.5 in the different scenario, which is realistic in terms of size (see definition above). To compare this model to others, we then define the parameters $\alpha_{homophily}$ and $\alpha_{triadic}$ that verify:

\begin{equation}
\hat{R}_H (\pi_{contact}, \alpha_{repetition}) = R_H (\pi_{contact} ,\alpha_{homophily}) = R_H (\pi_{contact},\alpha_{triadic})
\end{equation}

using a standard optimisation algorithm.  The average parameters across simulations for the different network scenarios are  $\alpha_{triadic}$=0.75 and $\alpha_{homophily}$=17.6. While that latter parameter appears large, note that the associated statistic $s_{homophily}$ ranges from 0 to 1, with most realised values close to 1. 

\subsection*{Parametrisation of the different simulations}

Unless otherwise noted, all simulations use $\pi_{contact}$=0.5 except for the null model, which uses $\pi_{contact}$=1. In all simulations except the ones that vary the infectiousness, $\pi_{infection}$=0.8. Unless otherwise noted, $T_{exposure}=1n$ and $T_{infection}=4n$. Given the substantial computational burden involved in conducting the simulations, 48 repetitions were run for networks with $n$ $\leq$ 1000, with 40 for larger networks. Experiments varying  $T_{exposure}$ and $\pi_{infection}$ used 24 repetitions.

For the experiments that vary the structure of the underlying network and the network size, the parameters that guide the stochastic network creation are presented in Table \ref{tables1}. Descriptive statistics of these networks are presented in Table \ref{tables2}. The underlying networks that are used in the other variation experiments are generated according to the parameters denoted `1: Baseline' in Table \ref{tables1}-\ref{tables2}. The four experiments that vary the time individuals are in the `exposed' state before becoming `infectious' use values for $T_{exposure}$ of 0, 1$n$, 2$n$, 3$n$ and 4$n$. The four experiments that vary the infectiousness of the disease use values for $\pi_{infection}$ of 0.55, 0.65, 0.8, and 0.95.

The experiment that used geography as the basis of the homophily strategy was created according to the `1: Baseline' parameters but used the Euclidean distance in geographic placement as the basis for choosing interaction partners in the homophily strategy. The two experiments on multidimensional homophily used underlying networks created following the `1: Baseline' parameters, with the exception that instead of one homophilous attribute, two attributes were defined and the number of ties created according to the homophily parameter was split evenly between the two dimensions. The homophily strategy used for the simulated infection curves in the two scenarios differs in the sense that in the first, individuals interact according to minimising the absolute difference in both attributes. In the second scenario, only the first attribute was used as the basis of the homophily strategy and the second attribute was ignored.

For the experiments using mixed strategies, the probability of partner choice $p(i \rightarrow j)$ can depend on a vector of statistics and parameters \citep{stadtfeld2017}. The entropy based on a set parameter vector was used to calibrate the parameter for the homophily and triadic closure strategy as comparison cases. Parameter choices rely on experimentation to result in similar entropy values as when using single strategies. For the mixed strategy of repetition and homophily, the parameters were set to $\alpha_{homophily}$=0.7  and $\alpha_{repetition}$=1.6. For the mixed strategy of repetition and triadic closure, the parameters were set to $\alpha_{triadic}$=0.35 and $\alpha_{repetition}$=1.6. For the mixed strategy of homophily and triadic closure, the parameters were set to $\alpha_{homophily}$=6 and $\alpha_{triadic}$=0.35. For the mixed strategy incorporating all three, parameters were set to $\alpha_{homophily}$=4, $\alpha_{triadic}$=0.3, and $\alpha_{repetition}$=1.2.

The simulated average infection curves for all experiments can be found in Figures \ref{figures1}-\ref{figures7}. Descriptive results for the simulations in terms of delay of peak, height of peak and total number infected at the end of the simulation are presented in Table \ref{tables3}. Note that the descriptive statistics in this table present the averages of characteristics of the repetitions of the simulated infection curves, which are not the same as the characteristics of the average infection curves as presented in the supplementary figures.\\

\noindent
\textbf{Acknowledgements:} PB, JBM, CR, RK and MCM are supported by The Leverhulme Trust, Leverhulme Centre for Demographic Science. CR is supported by the British Academy. MCM is supported by ERC Advanced Grant CHRONO (835079). The authors would like to thank Mark Verhagen, Valentina Rotondi, and Liliana Andriano for feedback.\\

\noindent
\textbf{Author contributions:} PB, MH, and IJR conceptualised the study; PB and MH contributed methodology, implementation and performed analyses; PB, MH, IJR, and MCM wrote initial manuscript and provided visualisation; all authors (PB, MH, IJR, JBD, CR, RK, and MCM) discussed research design and reviewed and edited manuscript. \\

\noindent
\textbf{Competing interests}: The authors declare no competing interests. \\

\bibliography{postlockdownn}\label{sec:Bibliography}
\newpage

\renewcommand{\thefigure}{S\arabic{figure}}
\setcounter{figure}{0}

\begin{figure}[!t]
\centering
\includegraphics[width=1\textwidth]{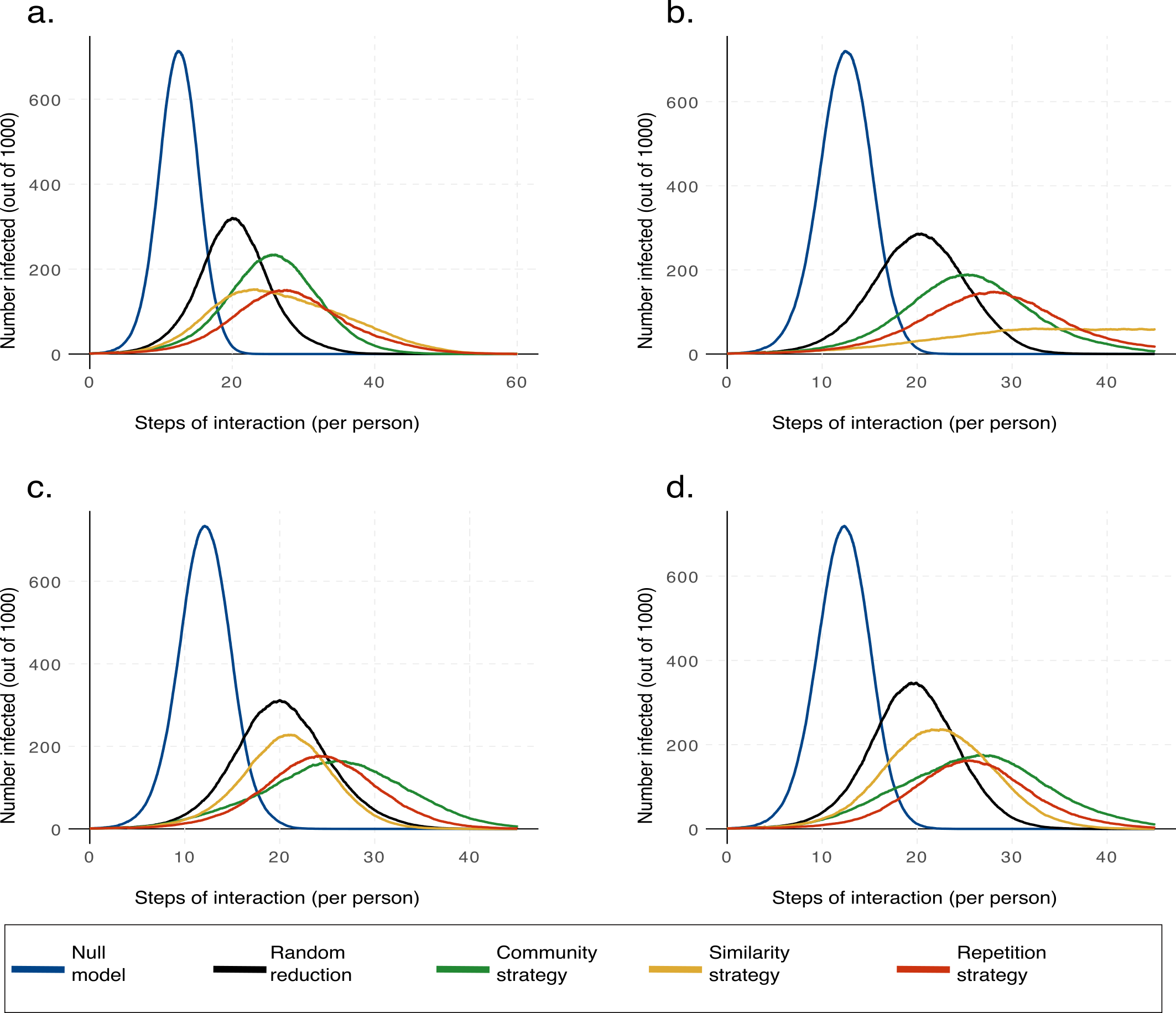}
\caption{Curves compare 4 contact reduction strategies to the null model of no social distancing, as described in the main text. (A) Reference model with standard operationalisation of homophily; (B) model with homophily based on geographic proximity; (C) underlying network model with homophily based on two dimensions, interaction strategy minimises the overall difference along both attributes; (D) underlying network model with homophily based on two dimensions, interaction strategy minimises the difference only on the first attribute.}\label{figures1}
\end{figure}
\newpage

\begin{figure}[!t]
\centering
\includegraphics[width=1\textwidth]{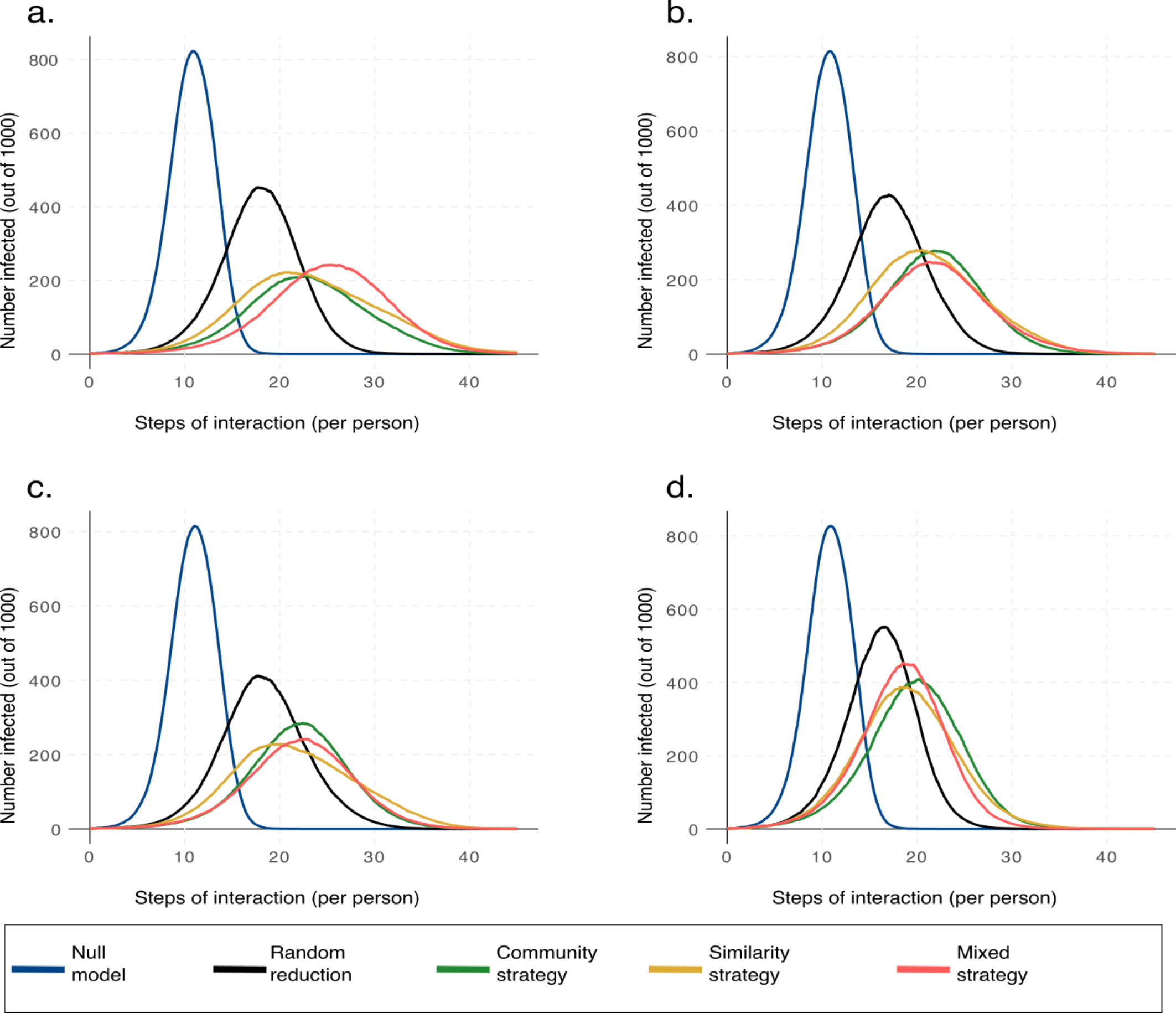}
\caption{Curves compare 4 contact reduction strategies to the null model of no social distancing, as described in the main text. (A) Mixed strategy of repetition and triadic closure; (B) mixed strategy of repetition and homophily; (C) mixed strategy of repetition, homophily, and triadic closure; (D) mixed strategy of homophily and triadic closure.}\label{figures2}
\end{figure}
\newpage

\begin{figure}[!t]
\centering
\includegraphics[width=1\textwidth]{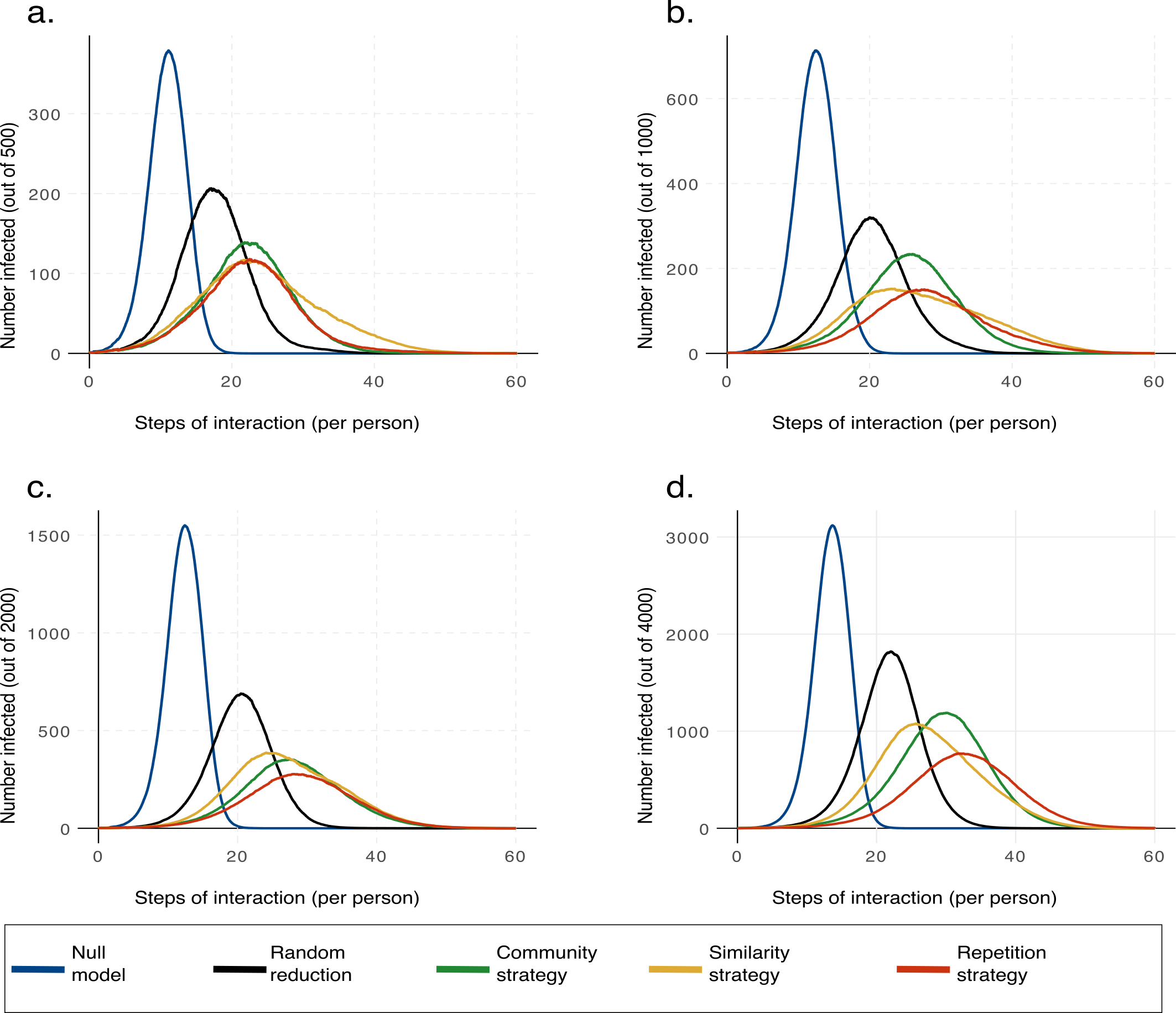}
\caption{Curves compare 4 contact reduction strategies to the null model of no social distancing, as described in the main text. (A) 500 actors; (B) 1000 actors; (C) 2000 actors; (D) 4000 actors.}\label{figures3}
\end{figure}
\newpage

\begin{figure}[!t]
\centering
\includegraphics[width=1\textwidth]{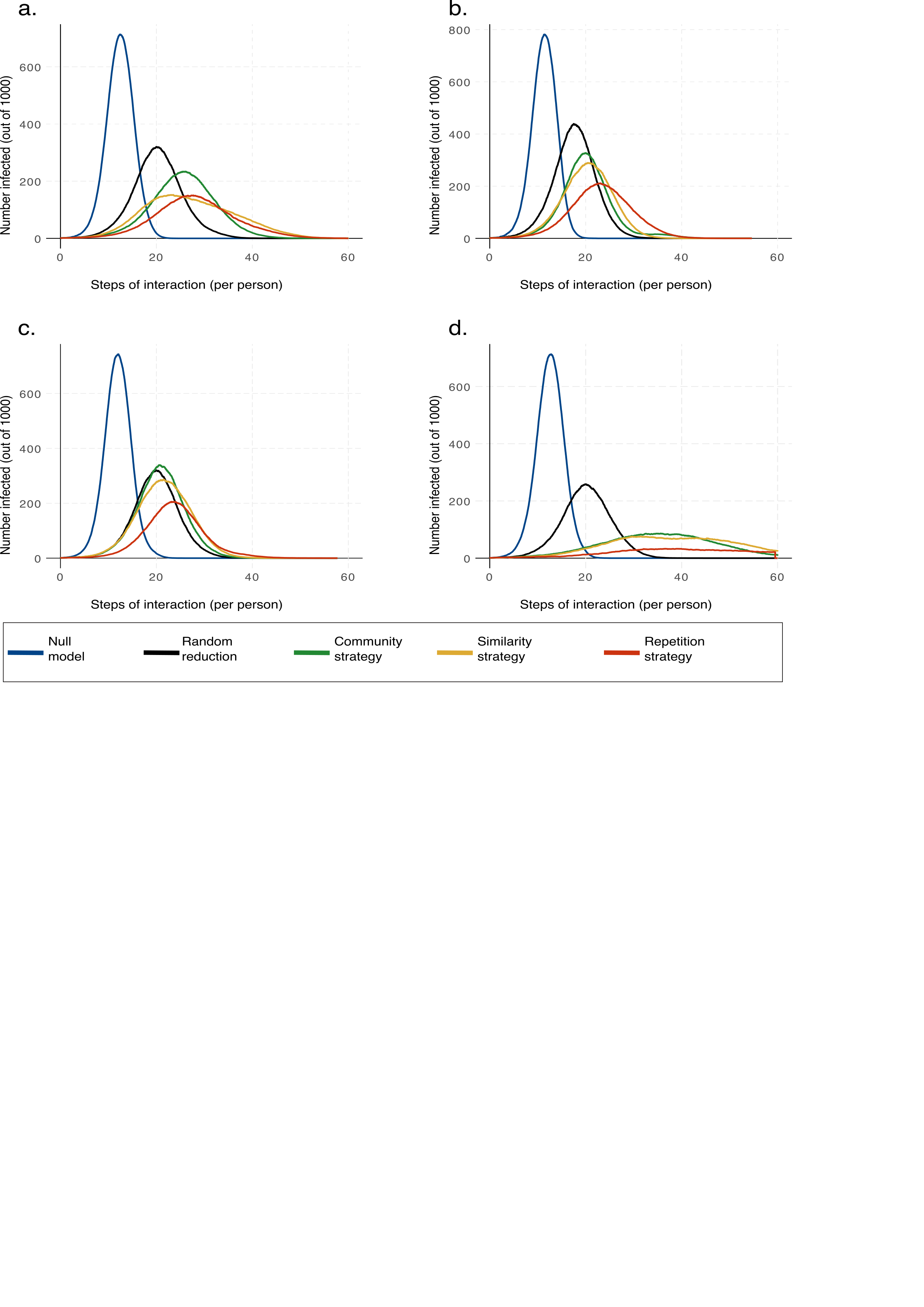}
\caption{Curves compare 4 contact reduction strategies to the null model of no social distancing, as described in the main text. Names refer to the parametrisation given in Table S1. (A) baseline scenario; (B) random network; (C) higher degree; (D) lower degree.}\label{figures4}
\end{figure}
\newpage

\begin{figure}[!t]
\centering
\includegraphics[width=1\textwidth]{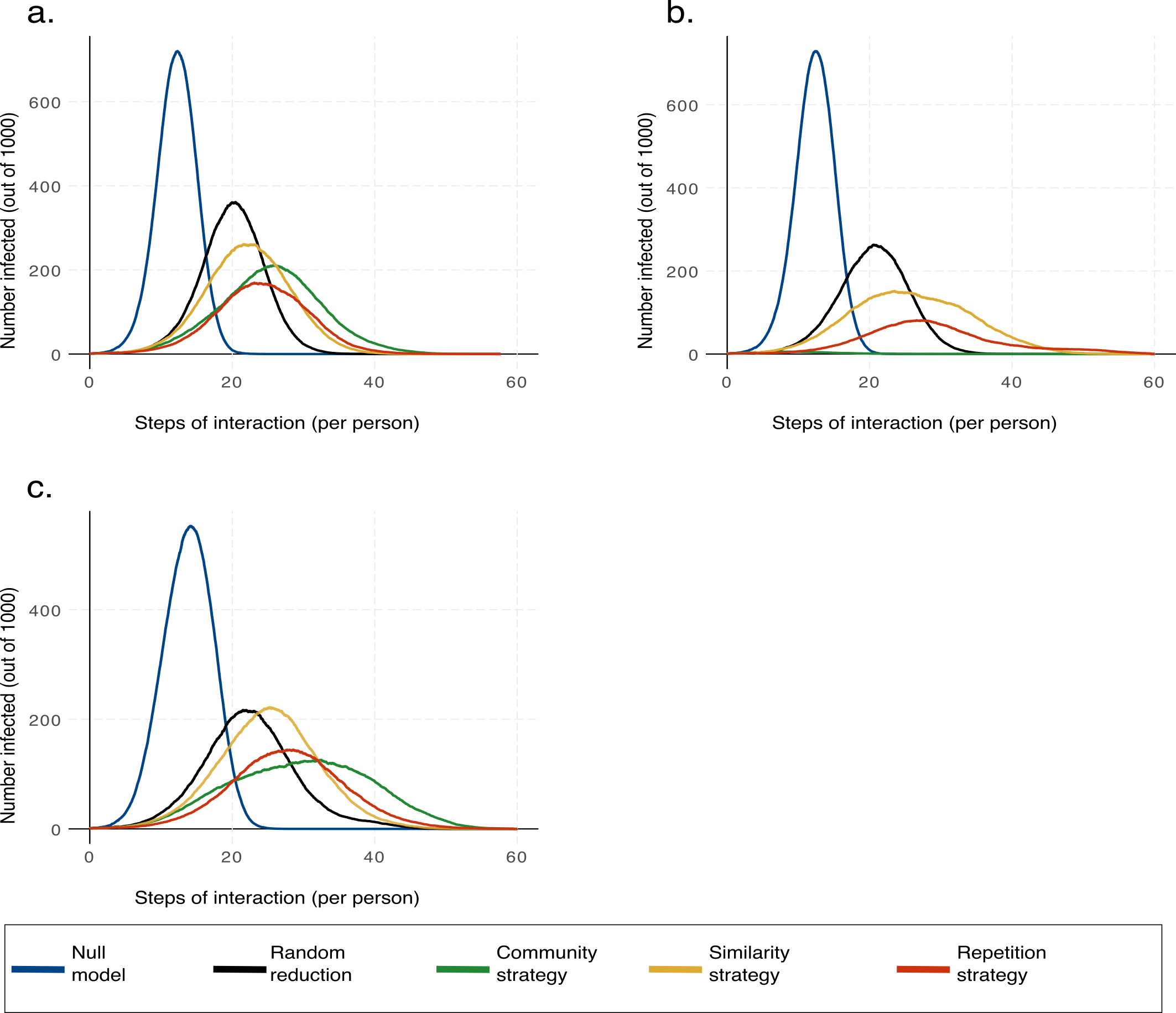}
\caption{Curves compare 4 contact reduction strategies to the null model of no social distancing, as described in the main text. Names refer to the parametrisation given in Table S1. (A) no groups; (B) no geography; (C) small world-ish.}\label{figures5}
\end{figure}
\newpage

\begin{figure}[!t]
\centering
\includegraphics[width=1\textwidth]{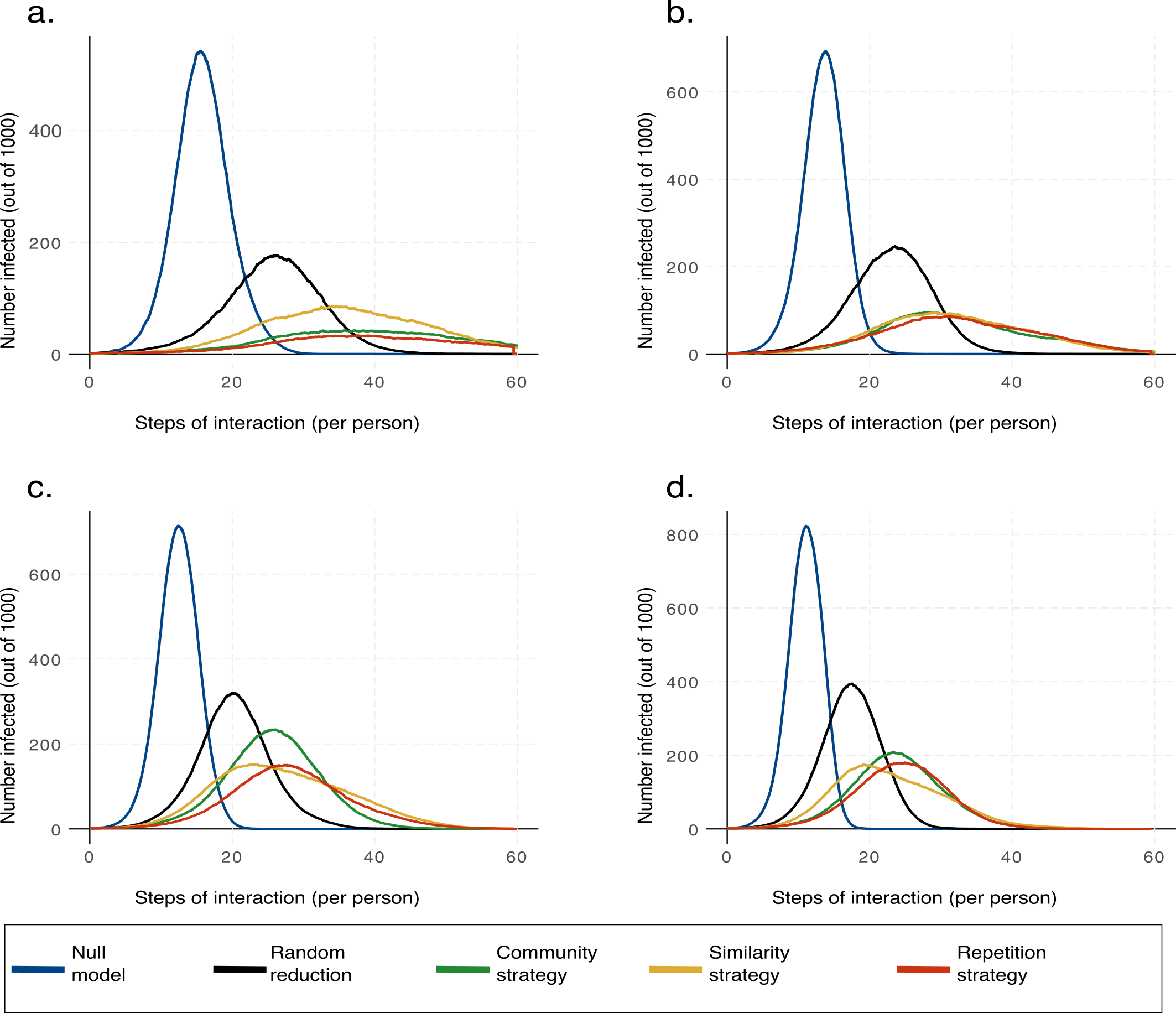}
\caption{Curves compare 4 contact reduction strategies to the null model of no social distancing, as described in the main text. (A) $\pi_{infection}$=0.55; (B) $\pi_{infection}$=0.65; (C) $\pi_{infection}$=0.8; (D)$\pi_{infection}$=0.95.}\label{figures6}
\end{figure}
\newpage

\begin{figure}[!t]
\centering
\includegraphics[width=1\textwidth]{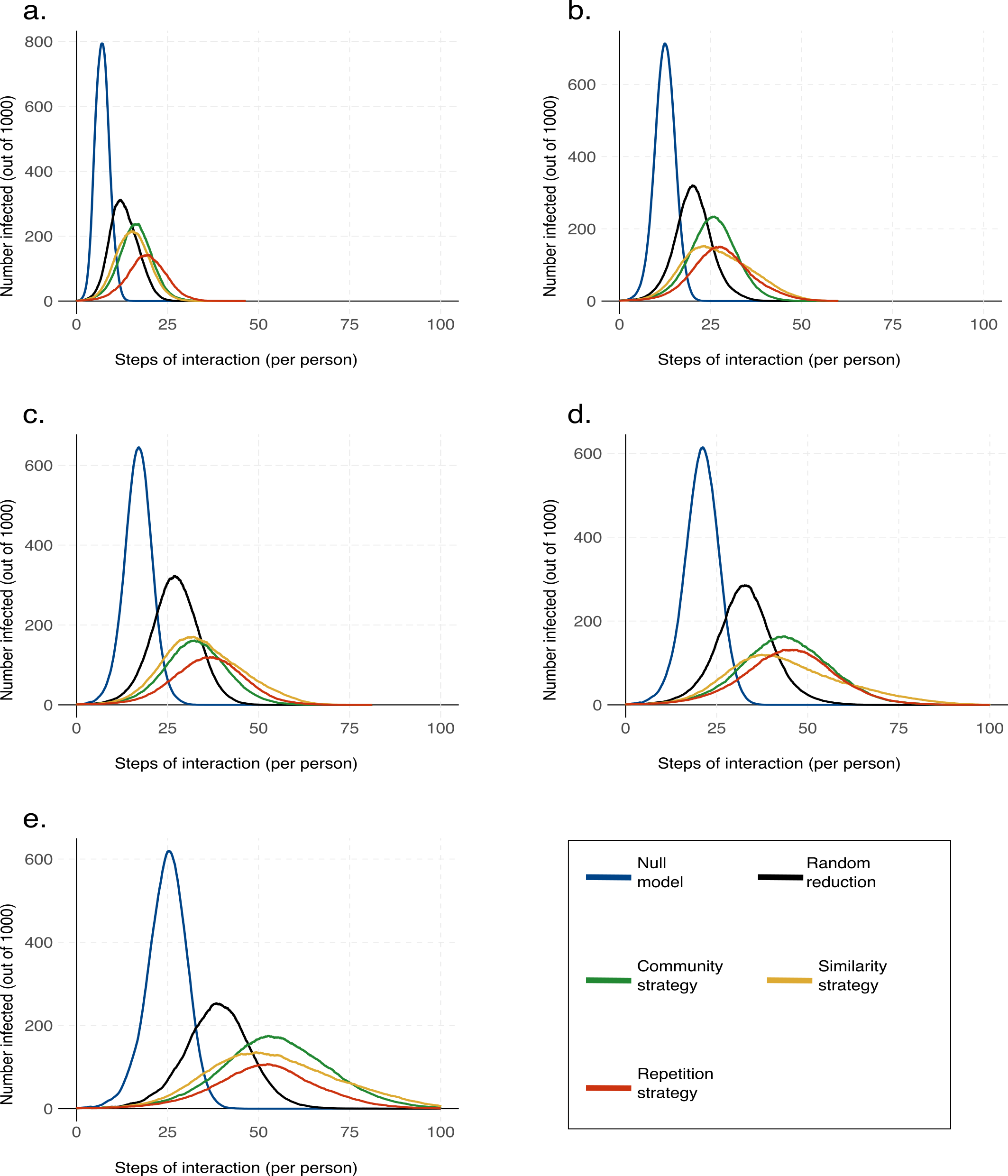}
\caption{Curves compare 4 contact reduction strategies to the null model of no social distancing, as described in the main text. (A) $T_{exposed}$=0; (B) $T_{exposed}$=1n; (C) $T_{exposed}$=2n; (D) $T_{exposed}$=3n; (E) $T_{exposed}$=4n.}\label{figures7}
\end{figure}
\clearpage
\newpage

\renewcommand{\thetable}{S\arabic{table}}
\setcounter{table}{0}

\newgeometry{margin=2cm} 
\begin{landscape}
\begin{table}[!h]
\begin{center}
\begin{tabular}{rcccccccccccccc}\toprule
\multicolumn{5}{l}{Scenario}             & nActors & d\_geo\_min & d\_geo\_max & g\_geo & m\_groups & g\_groups & d\_hom\_min & d\_hom\_max & w\_hom. & z\_random \\ \midrule
\multicolumn{5}{l}{1: Baseline scenario} & 1000    & 4           & 12          & 0.3    & 8         & 0.9       & 4           & 12          & 2       & 0.5       \\
\multicolumn{5}{l}{2: Higher degree}     & 1000    & 8           & 24          & 0.3    & 16        & 0.9       & 8           & 24          & 2       & 1         \\
\multicolumn{5}{l}{3: Lower degree}      & 1000    & 2           & 6           & 0.3    & 4         & 0.9       & 2           & 6           & 2       & 0.25      \\
\multicolumn{5}{l}{4: No groups}         & 1000    & 10          & 30          & 0.3    & 0         & NA        & 5           & 15          & 2       & 0.5       \\
\multicolumn{5}{l}{5: No geography}      & 1000    & 0           & 0           & NA     & 20        & 0.9       & 5           & 15          & 2       & 0.5       \\
\multicolumn{5}{l}{6: Random net.}       & 1000    & 0           & 0           & NA     & 0         & NA        & 0           & 0           & NA      & 32        \\
\multicolumn{5}{l}{7: Geography}         & 1000    & 15          & 45          & 0.3    & 0         & NA        & 0           & 0           & NA      & 0.5       \\
\multicolumn{5}{l}{8: 500 Actors}        & 500     & 4           & 12          & 0.3    & 8         & 0.9       & 4           & 12          & 2       & 0.5       \\
\multicolumn{5}{l}{9: 1000 Actors}       & 1000    & 4           & 12          & 0.3    & 8         & 0.9       & 4           & 12          & 2       & 0.5       \\
\multicolumn{5}{l}{10: 2000 Actors}      & 2000    & 4           & 12          & 0.3    & 8         & 0.9       & 4           & 12          & 2       & 0.5       \\
\multicolumn{5}{l}{11: 4000 Actors}      & 4000    & 4           & 12          & 0.3    & 8         & 0.9       & 4           & 12          & 2       & 0.5\\ \bottomrule
\end{tabular}
\caption{Parameters used in the stochastic generation of underlying networks. Full description of procedure is described in the Methods section in the main text.}\label{tables1}
\end{center}
\end{table}
\end{landscape}
\restoregeometry
\newpage

\begin{table}[]
\begin{center}
\begin{tabular}{lcccccc}\toprule
Scenario                 & n    & deg.  & clus. & av. 				path & dia. & hom. \\ \midrule
1: 				Baseline scenario & 1000 & 38.4  & 0.11  & 2.23         & 3    & 1.08 \\
2: 				Higher degree     & 1000 & 75.9  & 0.14  & 1.93         & 3    & 1.08 \\
3: 				Lower degree      & 1000 & 19.4  & 0.09  & 2.69         & 4    & 1.08 \\
4: 				No groups         & 1000 & 55.4  & 0.16  & 2.07         & 3    & 1.07 \\
5: 				No geography      & 1000 & 40.2  & 0.26  & 2.24         & 3    & 1.09 \\
6: 				Random net.       & 1000 & 62    & 0.06  & 1.96         & 3    & 1    \\
7: 				Small world-ish   & 1000 & 53.9  & 0.3   & 2.57         & 4    & 1    \\
8: 				500 Actors        & 500  & 38.11 & 0.14  & 2            & 3    & 1.08 \\
9: 				1000 Actors       & 1000 & 38.4  & 0.11  & 2.23         & 3    & 1.08 \\
10: 				2000 Actors      & 2000 & 38.72 & 0.09  & 2.49         & 3.4  & 1.08 \\
11: 				4000 Actors      & 4000 & 38.85 & 0.08  & 2.7          & 4    & 1.08\\ \bottomrule
\end{tabular}
\caption{\textbf{Characteristics of the networks created under different scenarios}. Descriptive statistics are averaged over 40-48 simulations. Notes: n: number of actors; deg.: average degree / number of connections per actor; clus.: clustering coefficient / proportion of closed triads over possibly closed triads; av. Path: average network distance between pairs of nodes; dia.: diameter / maximum distance in between nodes in the network; hom.: average similarity of interaction partners divided by average similarity among all actors.}\label{tables2}
\end{center}
\end{table}
\newpage
%\newgeometry{margin=1.3cm} % modify this if you need even more space
\begin{table}[]\label{tables3}
\begin{scriptsize}
\begin{tabular}{llccc}
\multicolumn{5}{l}{Variation: \textbf{network structure}}                     \\ \midrule
Scenario                  & Strategy         & Delay & Peak & Inf. \\ \midrule
1: 				Baseline scenario  & Random           & 1.30  & 0.50 & 75\% \\
                          & Triads           & 1.66  & 0.35 & 73\% \\
                          & Homophily        & 1.50  & 0.23 & 71\% \\
                          & Repetition       & 1.59  & 0.24 & 57\% \\
2: 				Higher degree      & Random           & 1.25  & 0.48 & 71\% \\
                          & Triads           & 1.49  & 0.48 & 79\% \\
                          & Homophily        & 1.52  & 0.44 & 80\% \\
                          & Repetition       & 1.24  & 0.31 & 56\% \\
3: 				Lower degree       & Random           & 1.06  & 0.40 & 62\% \\
                          & Triads           & 1.85  & 0.14 & 51\% \\
                          & Homophily        & 1.88  & 0.12 & 52\% \\
                          & Repetition       & 2.00  & 0.06 & 24\% \\
4: 				No groups          & Random           & 1.34  & 0.52 & 77\% \\
                          & Triads           & 1.68  & 0.29 & 72\% \\
                          & Homophily        & 1.46  & 0.38 & 77\% \\
                          & Repetition       & 1.24  & 0.27 & 54\% \\
5: 				No geography       & Random           & 1.12  & 0.41 & 63\% \\
                          & Triads           & 0.71  & 0.01 & 2\%  \\
                          & Homophily        & 1.38  & 0.22 & 65\% \\
                          & Repetition       & 0.99  & 0.14 & 32\% \\
6: 				Random net.        & Random           & 1.46  & 0.59 & 90\% \\
                          & Triads           & 1.38  & 0.45 & 72\% \\
                          & Homophily        & 1.39  & 0.41 & 73\% \\
                          & Repetition       & 1.46  & 0.32 & 63\% \\
7: 				Small world-ish    & Random           & 1.11  & 0.43 & 65\% \\
                          & Triads           & 1.61  & 0.25 & 66\% \\
                          & Homophily        & 1.44  & 0.45 & 74\% \\
                          & Repetition       & 1.30  & 0.29 & 56\% \\ \bottomrule
                          &                  &       &      &      \\
\multicolumn{5}{l}{Variation: \textbf{network size}}                     \\ \midrule
8: 				500 Actors         & Random           & 1.59  & 0.62 & 94\% \\
                          & Triads           & 1.83  & 0.41 & 81\% \\
                          & Homophily        & 2.09  & 0.33 & 93\% \\
                          & Repetition       & 1.87  & 0.34 & 75\% \\
9: 				1000 Actors        & Random           & 1.30  & 0.50 & 75\% \\
                          & Triads           & 1.66  & 0.35 & 73\% \\
                          & Homophily        & 1.50  & 0.23 & 71\% \\
                          & Repetition       & 1.59  & 0.24 & 57\% \\
10: 				2000 Actors       & Random           & 1.30  & 0.49 & 74\% \\
                          & Triads           & 1.49  & 0.28 & 60\% \\
                          & Homophily        & 1.60  & 0.26 & 73\% \\
                          & Repetition       & 1.47  & 0.21 & 52\% \\
11: 				4000 Actors       & Random           & 1.63  & 0.63 & 95\% \\
                          & Triads           & 2.20  & 0.42 & 93\% \\
                          & Homophily        & 1.90  & 0.36 & 94\% \\
                          & Repetition       & 2.08  & 0.29 & 71\% \\ \bottomrule
                          &                  &       &      &      \\
\multicolumn{5}{l}{Variation: \textbf{infectiousness}}                   \\ \midrule
12: 				$p_{infect}$=0.55    & Random           & 1.11  & 0.32 & 57\% \\
                          & Triads           & 0.96  & 0.10 & 29\% \\
                          & Homophily        & 1.25  & 0.16 & 49\% \\
                          & Repetition       & 1.25  & 0.07 & 22\% \\
13: 				$p_{infect}$=0.65    & Random           & 1.29  & 0.42 & 68\% \\
                          & Triads           & 1.34  & 0.19 & 47\% \\
                          & Homophily        & 1.21  & 0.16 & 48\% \\
                          & Repetition       & 1.65  & 0.17 & 46\% \\
14: 				$p_{infect}$=0.8     & Random           & 1.30  & 0.50 & 75\% \\
                          & Triads           & 1.66  & 0.35 & 73\% \\
                          & Homophily        & 1.50  & 0.23 & 71\% \\
                          & Repetition       & 1.59  & 0.24 & 57\% \\
15: 				$p_{infect}$=0.95    & Random           & 1.32  & 0.57 & 80\% \\
                          & Triads           & 1.44  & 0.33 & 62\% \\
                          & Homophily        & 1.33  & 0.24 & 63\% \\
                          & Repetition       & 1.48  & 0.27 & 57\% \\ \bottomrule
                          &                  &       &      &    
\end{tabular}
 \begin{tabular}[!t]{llccc}
\multicolumn{5}{l}{Variation: \textbf{$T_{exposure}$}}                     \\ \midrule
Scenario                  & Strategy         & Delay & Peak & Inf. \\ \midrule
16: 				$T_{exposure}$ = 0   & Random           & 1.44  & 0.50 & 74\% \\
                          & Triads           & 1.69  & 0.33 & 65\% \\
                          & Homophily        & 1.48  & 0.27 & 63\% \\
                          & Repetition       & 1.52  & 0.21 & 46\% \\
17: 				$T_{exposure}$ = 1   & Random           & 1.30  & 0.50 & 75\% \\
                          & Triads           & 1.66  & 0.35 & 73\% \\
                          & Homophily        & 1.50  & 0.23 & 71\% \\
                          & Repetition       & 1.59  & 0.24 & 57\% \\
18: 				$T_{exposure}$ = 2   & Random           & 1.40  & 0.54 & 84\% \\
                          & Triads           & 1.20  & 0.26 & 56\% \\
                          & Homophily        & 1.54  & 0.26 & 74\% \\
                          & Repetition       & 1.24  & 0.19 & 47\% \\
19: 				$T_{exposure}$ = 3   & Random           & 1.22  & 0.47 & 73\% \\
                          & Triads           & 1.58  & 0.32 & 68\% \\
                          & Homophily        & 1.24  & 0.21 & 62\% \\
                          & Repetition       & 1.49  & 0.23 & 57\% \\
20: 				$T_{exposure}$ = 4   & Random           & 1.11  & 0.43 & 68\% \\
                          & Triads           & 1.85  & 0.34 & 78\% \\
                          & Homophily        & 1.50  & 0.23 & 73\% \\
                          & Repetition       & 1.11  & 0.19 & 44\% \\ \bottomrule
                          &                  &       &      &      \\
\multicolumn{5}{l}{Variation: \textbf{mixed strategies}}                 \\ \midrule
21: 				Homo. + Triad.    & Random           & 1.52  & 0.69 & 97\% \\
                          & Triads           & 1.86  & 0.55 & 96\% \\
                          & Homophily        & 1.73  & 0.49 & 97\% \\
                          & Mixed strat. & 1.75  & 0.58 & 96\% \\
22: 				Homo. + Rep.      & Random           & 1.39  & 0.60 & 84\% \\
                          & Triads           & 1.51  & 0.40 & 70\% \\
                          & Homophily        & 1.60  & 0.37 & 82\% \\
                          & Mixed strat. & 1.52  & 0.36 & 68\% \\
23: 				Rep. + Triad      & Random           & 1.53  & 0.63 & 89\% \\
                          & Triads           & 1.47  & 0.34 & 64\% \\
                          & Homophily        & 1.67  & 0.31 & 80\% \\
                          & Mixed strat. & 1.92  & 0.35 & 74\% \\
24: 				Ho. + Re. + Tr.   & Random           & 1.53  & 0.63 & 88\% \\
                          & Triads           & 1.51  & 0.39 & 70\% \\
                          & Homophily        & 1.43  & 0.32 & 74\% \\
                          & Mixed strat. & 1.39  & 0.34 & 64\% \\ \bottomrule
                          &                  &       &      &      \\
\multicolumn{5}{l}{Variation: \textbf{operationalisation homophily}}     \\ \midrule
25: 				Normal homo.      & Random           & 1.30  & 0.50 & 75\% \\
                          & Triads           & 1.66  & 0.35 & 73\% \\
                          & Homophily        & 1.50  & 0.23 & 71\% \\
                          & Repetition       & 1.59  & 0.24 & 57\% \\
26: 				Geogr. homo.      & Random           & 1.20  & 0.46 & 69\% \\
                          & Triads           & 1.40  & 0.30 & 62\% \\
                          & Homophily        & 2.19  & 0.11 & 50\% \\
                          & Repetition       & 1.55  & 0.23 & 54\% \\
27: 				2-Dim. abs. diff. & Random           & 1.31  & 0.50 & 75\% \\
                          & Triads           & 1.47  & 0.25 & 61\% \\
                          & Homophily        & 1.00  & 0.31 & 53\% \\
                          & Repetition       & 1.22  & 0.26 & 52\% \\
28: 				2-Dim. only 1st   & Random           & 1.32  & 0.52 & 78\% \\
                          & Triads           & 1.63  & 0.27 & 67\% \\
                          & Homophily        & 1.33  & 0.36 & 68\% \\
                          & Repetition       & 1.19  & 0.25 & 50\% \\ \bottomrule
                          &                  &       &      &      \\
\multicolumn{5}{p{9cm}}{\textbf{Table S3: Characteristics of average infection curves for different strategies.} All entries denoting averaged results of simulations are relative to the null model of no contact reduction (blue line in Fig. 3). Delay: delay of the peak of the infection curve compared to the null model; Peak: height of the peak of the infection curve compared to the null model; Inf.: proportion of the population infected compared to the null model.}\\
                          &                  &       &      &      \\
\end{tabular}\label{tables3}
\end{scriptsize}
\fakecaption\label{tables3}
\end{table}
\end{document}